\documentclass{INTERSPEECH2023}

\interspeechcameraready

\title{Speech inpainting: Context-based speech synthesis guided by video}
\name{Juan F. Montesinos$^1$,
Daniel Michelsanti$^{2,3}$, Gloria Haro$^1$, Zheng-Hua Tan$^2$, Jesper Jensen$^{2,3}$}
\address{$^1$Universitat Pompeu Fabra, Department of Information and Communication Technologies, Spain\\
         $^2$Aalborg University, Department of Electronic Systems, Denmark\\
         $^3$Oticon A/S, Denmark
         }
\email{\{juanfelipe.montesinos,gloria.haro\}@upf \{danmi,zt,jje\}@es.aau.dk}

\begin{document}

\maketitle
 
\begin{abstract}
Audio and visual modalities are inherently connected in speech signals: lip movements and facial expressions are correlated with  speech sounds. This motivates studies that incorporate the visual modality to enhance an acoustic speech signal or even restore missing audio information. Specifically, this paper focuses on the problem of audio-visual speech inpainting, which is the task of synthesizing the speech in a corrupted audio segment in a way that it is consistent with the corresponding visual content and the uncorrupted  audio context. We present an  audio-visual transformer-based deep learning model that leverages visual cues that provide information about the content of the corrupted audio. It outperforms the previous state-of-the-art audio-visual model and audio-only baselines. We also show how visual features extracted with AV-HuBERT, %\cite{avhubert}, 
a large audio-visual transformer for speech recognition, are suitable for synthesizing speech. 
%Demos of our system are available at \url{https://anonymousarticle.github.io/avsi/}.
\end{abstract}

\noindent\textbf{Index Terms}:
speech, inpainting, audio-visual, transformer, multimodal, deep learning
\section{Introduction}
\label{sec:intro}
Speech is one of the most common multimodal events in our daily life. Thanks to the expansion of the Internet, we are exposed to a lot of speech signals from digital content as well: news, social networks, virtual meetings and video calls. Sometimes, the audio stream is corrupted due to, e.g., muted microphones, external noises or transmission losses. One solution is to estimate the lost audio information, saving content creators the time to re-make their videos or avoiding a speaker to repeat a sentence. The process of restoring the corrupted audio signal is known as \textit{audio inpainting} \cite{6020748}. Carrying out such a restoration for long segments of corrupted audio ($>$200ms) is not a simple task, as there is no prior information about the missing content. There are several ways to address the problem. From an audio-only (AO) perspective, the work in \cite{ebner2020audio} relies on a generative adversarial network approach to generate  realistic speech content for a gap size up to 500 ms. In \cite{chang2019deep}, an encoder-decoder architecture is used to inpaint the audio in both time-frequency and time domain for segments up to 250 ms. The works in \cite{marafioti2019context} and \cite{kegler2019deep} propose a similar idea operating only in the time-frequency domain for gaps up to 64 ms and 400 ms, respectively.

There are works using additional modalities, that are not affected by the acoustic noise, as cues to guide the inpainting process. This allows to inpaint larger gaps. For example, \cite{speechpainter} uses text to guide the inpainting process of audio gaps up to 1000 ms, relying on transformers and contrastive learning. In \cite{morrone2021audio}, video information is extracted from face landmarks to inpaint gaps up to 1600 ms using Bi-directional Long-Short Term Memory (Bi-LSTM)
units. 
The task of restoring a missing audio segment by leveraging the visual information of the speaker is known as audio-visual speech inpainting (AVSI). 
We present in this work an AVSI deep learning model which can restore long gaps of speech. In contrast to \cite{morrone2021audio}, we use high-level visual features useful for speech recognition and a multi-modal transformer that allows to establish  long range interactions across the audio and visual modalities, while being robust to potential misalignments among them. Moreover, the work in \cite{morrone2021audio} was limited to a constrained dataset of non natural speech \cite{grid}, while we train and test our model in a large-scale dataset of natural and unconstrained speech \cite{voxceleb} (in addition to \cite{grid}).

The contribution of this paper is two-fold: First, we propose a transformer architecture that analyzes a time-frequency representation of the corrupted audio signal and the corresponding uncorrupted visual information to synthesize intelligible speech even for a long corrupted audio segment, obtaining state-of-the-art results. Secondly, we show that speech inpainting can benefit from using high-level visual features extracted with the Audio-Visual HuBERT Network (AV-HuBERT) \cite{avhubert}, whose effectiveness for related tasks has previously been reported.

% The rest of this paper is structured as follows: ...

\section{Approach}

\subsection{Signal Model}\label{signal_model}
Let $x[t]$ be a discrete-time acoustic speech signal and 
$X=\{X(k,l); k=0,\dots,K-1 ; l=0,\dots,L-1\}$ %$X[k,l]$
be the corresponding  short-time Fourier transform (STFT), where $k$ and $l$ indicate  frequency and  time indices, respectively. Furthermore, let $\mathcal{A}\in \mathbb{R}^{K\times L}$ denote a magnitude spectrogram matrix defined from the element-wise absolute values of the elements in $X$ 
and $\mathcal{M} \in \mathbb{R}^{K\times L}$  a binary mask that provides the position of the corrupted region of the spectrogram \cite{morrone2021audio, paulino2020paco}). Then, the inpainted magnitude spectogram, $\mathcal{Q} \in \mathbb{R}^{K\times L}$, can be defined as
$\mathcal{Q}=\mathcal{M}\odot \mathcal{A} + (\textbf{1}-\mathcal{M})\odot\hat{\mathcal{A}},$
where $\odot$ indicates the element-wise product and $\hat{A} \in \mathbb{R}^{K\times L}$ denotes an estimated speech STFT magnitude matrix.  For the binary mask matrix $\mathcal{M}$, we assume that the $i$-th column consists of ones if the $i$-th column of $A$ is uncorrupted and zeros otherwise.

\subsection{Proposed Framework}\label{model}
AVSI leverages the video stream to improve speech inpainting, by providing  information about the acoustic speech content within the corrupted region. 
Our processing pipeline is divided into four different stages: feature extraction, multi-modal fusion, inpainting process and waveform reconstruction. The whole process is depicted in Fig. \ref{fig:model}.

\textbf{In the feature extraction stage}, we extract high-level visual features
% ($\mathcal{V} \in \mathbb{R}^{H\times W\times 3\times T}$)
using the AV-HuBERT's \cite{avhubert} video encoder, which processes the sequence of video frames using a ResNet \cite{resnet} followed by a transformer encoder to model the temporal dependencies. In addition, we use a simple multi-layer perceptron (MLP) with exponential linear unit (ELU) activation on top, leading to a signal $v\in \mathbb{R}^{D\times T}$, where $D$ is the dimensionality of our embeddings and T the amount of frames. In order to extract learned acoustic features, we use a similar MLP that takes as input the masked spectrogram $X\odot\mathcal{M}$, resulting in a signal $a \in \mathbb{R}^{D\times L}$ (see Fig.\ \ref{fig:model}). The aim behind this design is to process each audio frame independently, as we assume many audio frames can be corrupted. 

\textbf{In the multi-modal fusion stage}, the goal is to fuse the acoustic and visual features, learning the relationship between both. To do so, we rely on a six-block transformer encoder that ingests an audio-visual (AV) embedding. We construct the AV embedding by concatenating both modalities temporally, as in \cite{avhubert,rahimi2022reading,kadandale2022vocalist,chen2021audio}. Since the transformer is unaware of the position or the modality type of each element in the sequence, we sum a positional encoding ($pe$) that reflects the temporal sorting of the elements in the sequence \cite{montesinos2022vovit} and a modality encoding ($me$) that transmits whether each element is an acoustic or a visual feature \cite{chen2021audio}, obtaining:
%. The whole process is expressed in Eq. \ref{signals}:
\begin{equation}\label{signals}
%\begin{split}
    (a\, ; v) =  (pe_a+me_a+a \,; pe_v+me_v+v)
%\end{split}
\end{equation}
where $(\cdot\,;\cdot)$ denotes the concatenation of two sequences, resulting in an AV sequence $(a\, ; v) \in \mathbb{R}^{D\times (T+L)}$.  Alternatively, channel-wise concatenated AV embeddings can be used. Compared to the temporal concatenation, channel-wise concatenation would require an extra hyperparameter related to the number of features devoted to the visual and acoustic signals. Besides, due to the difference in the sampling rate, visual features should be upsampled to the temporal size of the audio ones.  On the other hand, we empirically found  that, in case of an out-of-sync AV stream, temporal concatenation results in predictions which are shifted in time, whereas in the channel-wise case, the system collapses and generates mumbling. An out-of-sync AV stream may occur due to software or hardware issues: codecs, latency, missing frames and it is frequent in low-quality videos. The robustness of the temporal concatenation to out-of-sync (i.e.\ misaligned) audio-visual pairs was also noticed in \cite{rahimi2022reading}. The downside effect is that the final sequence is larger, thus implying more computational cost.

\textbf{In the inpainting stage,} we use a seven-block transformer that processes the high-level features generated by the encoder to provide an estimate $\hat{\mathcal{A}}$ of the underlying uncorrupted speech magnitude spectogram. At this stage, the transformer's role is two-fold: It has to act as an auto-encoder, i.e.\ reconstruct the uncorrupted segment of the audio, and it has to inpaint the corrupted segment. 

\textbf{In the waveform reconstruction stage}, we estimate the phase of the underlying uncorrupted speech spectrogram using Local Weighted Sums (LWS)  \cite{lws} and then compute the  inverse STFT to recover the waveform, as done in \cite{morrone2021audio} (for a fair  comparison).%, although learning-based synthesizers such as \cite{} could be used).

\subsection{Training Loss, Data Pre-Processing and Model Setup}
We downsample the waveforms to 16 kHz. We compute the STFT with a hop size of 256, and a Hanning window of length 512. To process the video, we crop the mouth region, resizing the resulting frames to $96 \times 96$. Lastly, we extract the visual features as described in Section \ref{model}.

The transformer ingests 512-element embeddings across 8 heads. The dimensionality of the transformer's feed-forward layer is 1024. We use Gaussian error linear units (GELU) \cite{gelu} activation for the transformer and ELU \cite{elu} everywhere else. 
We train the model with a batch size of 10, a learning rate of $10^{-4}$ and the ADAM optimizer. As loss function we use a weighted mean absolute error (\textit{MAE}): 
\begin{equation} \label{eq:loss}%$$
\mathcal{L}(\mathcal{A}, \hat{\mathcal{A}}) = \alpha \, MAE(\hat{\mathcal{A}}^c , \mathcal{A}^c)+\beta \, MAE( \hat{\mathcal{A}}^u , \mathcal{A}^u),
\end{equation}%$$$ 
where $\alpha,\beta \ge 0$, and the superindices \textit{c} and \textit{u} denote corrupted and uncorrupted parts, respectively. We set $\alpha  > \beta$, so that  the network is forced to focus on the inpainting task, as it is much harder than the auto-encoding task (we use $\alpha=10$ and $\beta=1$).  %The $\alpha$ and $\beta$ coefficients are 10 and 1.

\begin{figure}[t]
    \centering
    \includegraphics[width=0.5\textwidth]{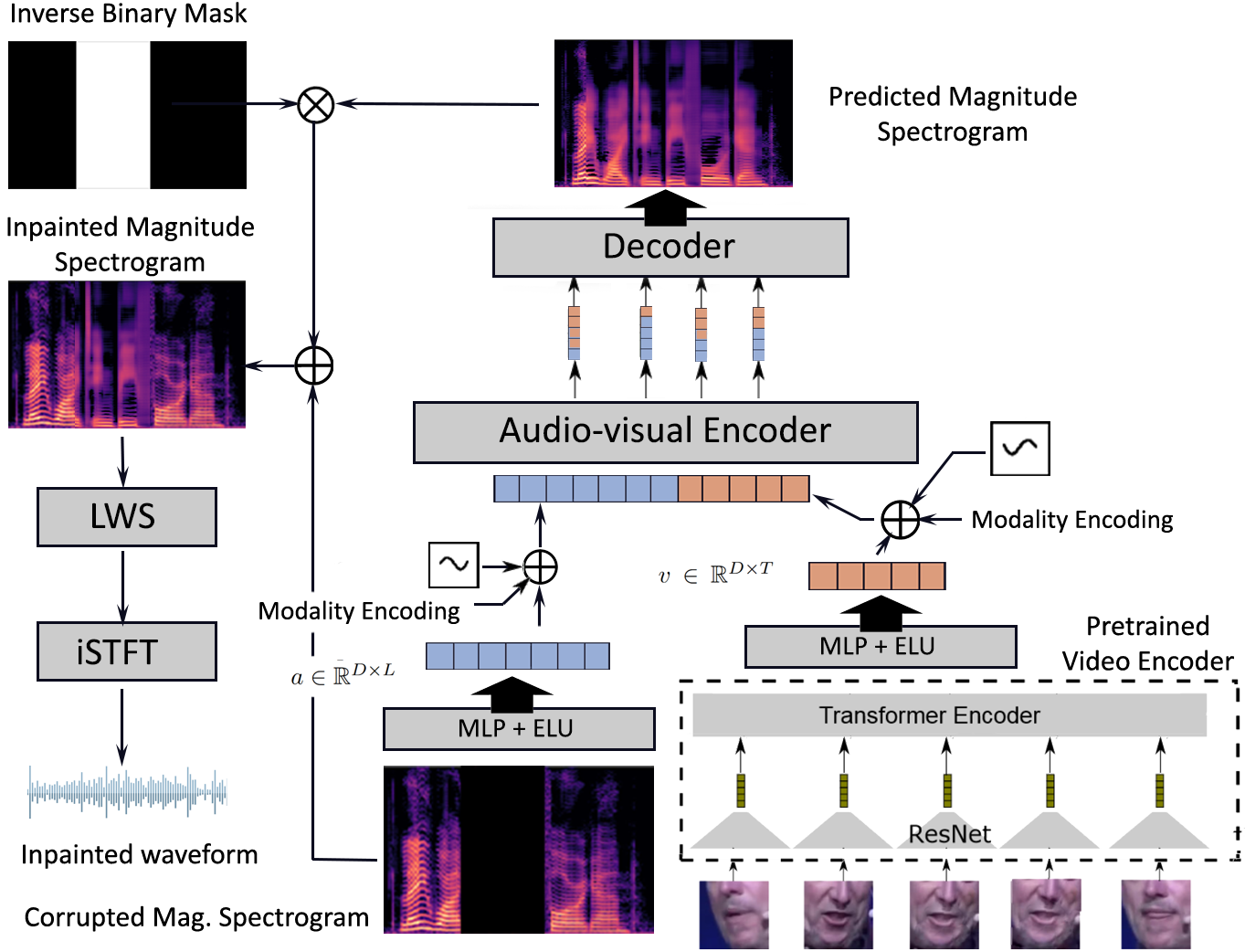}
    \caption{Proposed audio-visual model. The pre-trained video encoder corresponds to \cite{avhubert}.}
    \label{fig:model}
\end{figure}

 While training, we use the loss \eqref{eq:loss}, so the network predicts the whole spectrogram (both the corrupted and uncorrupted parts). At inference, once the spectrogram is predicted, we replace the known parts of the spectrogram via masking as shown in Figure \ref{fig:model}.

\begin{figure*}[ht]
    \centering
    \includegraphics[width=0.78\textwidth]{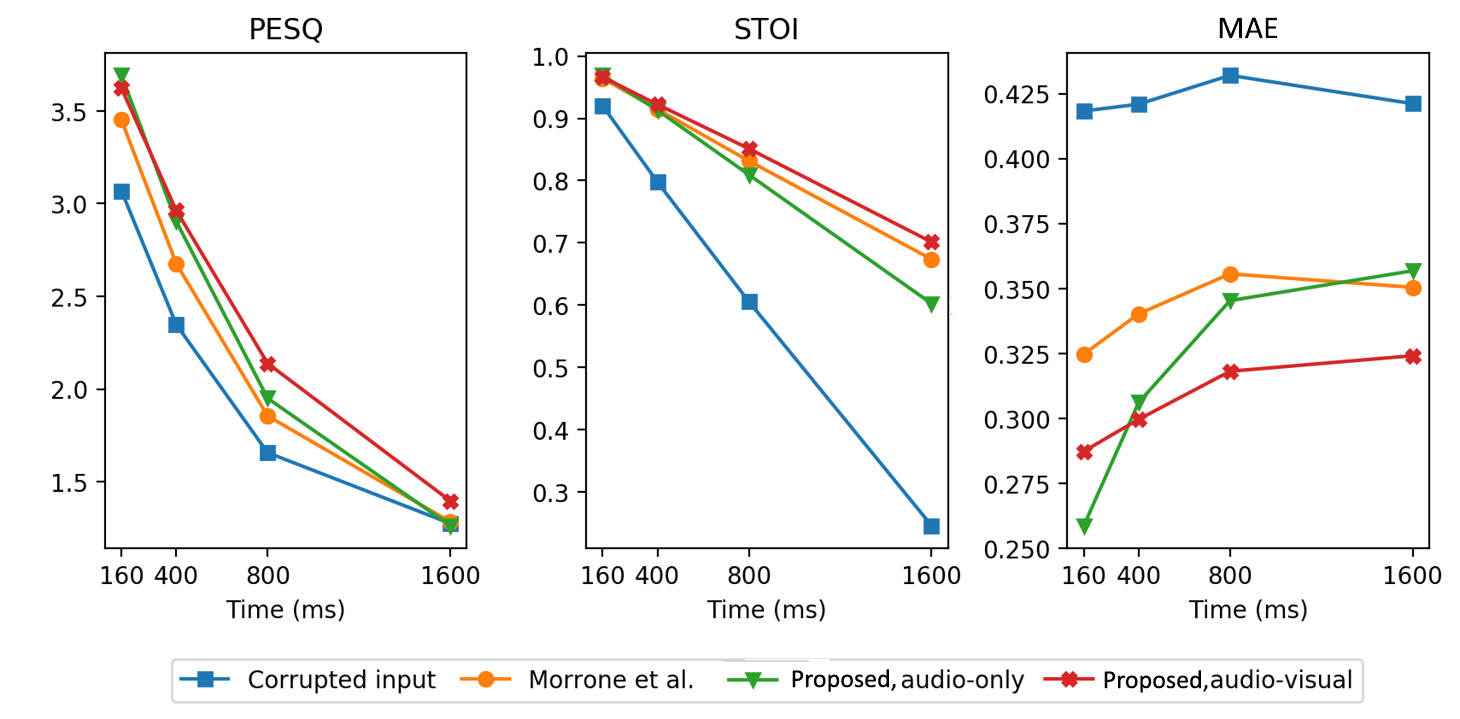}
    \caption{Comparison of performance vs corruption duration evaluated in the \textit{Grid Corpus} test set (see Sec. \ref{dataset}).}
    \label{fig:galaxy}
\end{figure*}

\section{Experiments} \label{expres}
\subsection{The Datasets} \label{dataset}
We train our model and the baselines using two different datasets: the \textit{Grid Corpus} \cite{grid} and the \textit{Voxceleb2} dataset \cite{voxceleb}.

The \textit{Grid Corpus} is a 30-hour AV dataset consisting of 33 speakers recorded in a controlled environment with a chroma screen as background, a frontal point of view, controlled lightning and a small vocabulary. 
Each video is 3 s long, recorded at 25 fps for the video and at 50 kHz for the audio. We split the dataset into training, validation and testing as in \cite{morrone2021audio}, for a fair comparison. 

The \textit{Voxceleb2} dataset is a large-scale dataset consisting of in-the-wild recordings of celebrities, which contains unconstrained natural vocabulary. Videos vary in duration and sampling rate. We trained the system with 2-second excerpts trimmed randomly. We select only those videos that are in English, the predominant language in the dataset, in order to discard biases in the results due to the language distribution. 

We corrupt the speech data with fullband temporal gaps of a duration between 160 and 1600 ms. During training, the corrupted segments are distributed randomly along each sample in a batch. During validation we apply the same logic so that the distribution of the validation set is as close as possible to the one of training. During testing, we run the system in 5 different setups: a random distribution of the gaps, as described before; corrupted segments with a gap of size 160 ms, 400 ms, 800 ms and 1600 ms.  The \textit{Grid} sentences typically include initial and trailing silence regions. When corrupting the speech signals, we ensure that the entire corrupted segment is located in the speech  active parts of the \textit{Grid} sentences (unlike \cite{morrone2021audio}, where corrupted segments were randomly chosen). 
%In contrast, in \cite{morrone2021audio}, the corrupted segment is randomly chosen and may include trailing silent parts which are easier to inpaint.

\subsection{Audio-Only and Audio-Visual Baselines}
We compare the proposed AVSI model against the previous state-of-the-art AV model, proposed in \cite{morrone2021audio}, and against the AO version of our model.
In the AV baseline \cite{morrone2021audio}, the authors propose a framework whose core is a stack of three Bi-LSTM layers fed with an AV signal. As acoustic features, they use normalized log magnitude spectrograms, while the visual features are landmark-based motion vectors. %based on a sequence of face landmarks. A sequence of face landmarks is a graphical representation of the motion and the spatial structure of the face. These are extracted by a deep neural network. 
In order to fuse the acoustic and the visual features via concatenation, they upsample the visual features to the sampling rate of the spectrogram. %Then, they compute the first temporal derivative of the landmarks to obtain motion vectors and they concatenate them to the spectrogram. 
%They minimize the MSE loss between the predicted log spectrogram of the corrupted segment and the ground-truth log spectrogram of the corrupted segment. 
They minimize the mean squared error of the predicted log magnitude spectrogram with respect to the ground-truth one in the corrupted segment.
Note that this is different from our setup, as we apply the loss on the whole predicted signal, not only in the corrupted segment. 

We also train our model in an AO setup, i.e., without visual information as input. This baseline permits to explore the benefits of using the additional modality of the video stream.

\begin{figure*}[t]
    \centering
    \includegraphics[width=0.8\textwidth]{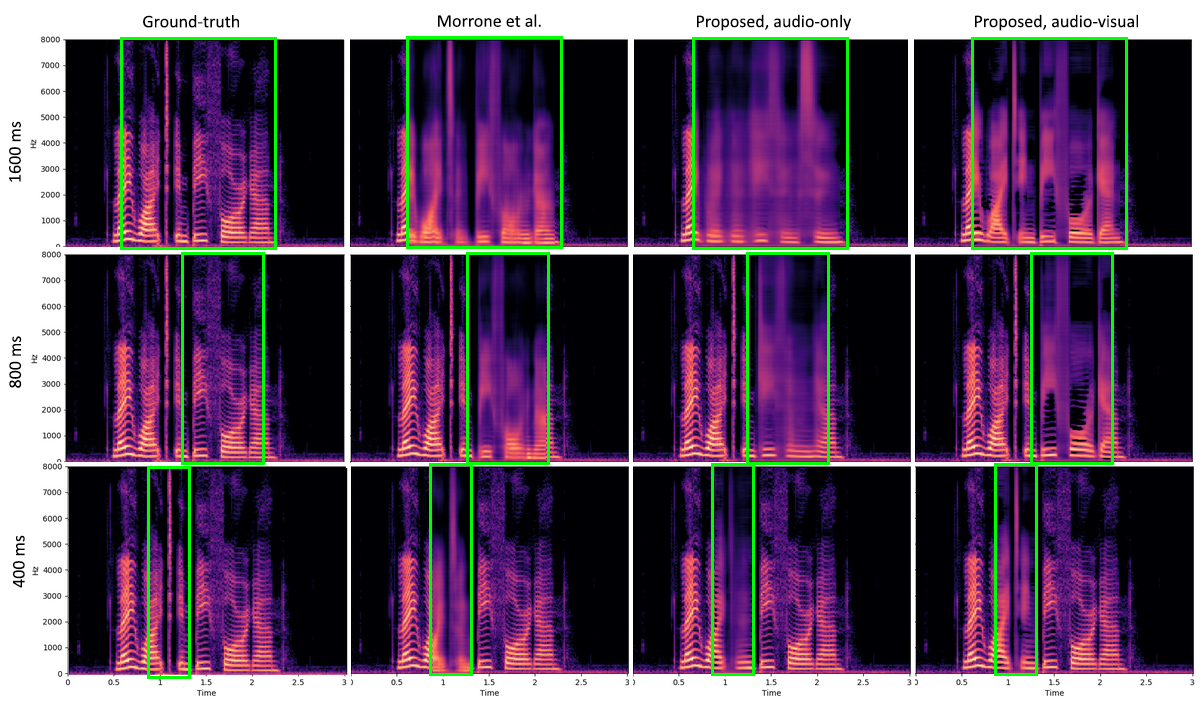}
    \caption{Sentence \textit{lwib4a} for speaker 34 in the \textit{Grid corpus} test set. Transcription:  \textit{"lay white in b four again"}. The region within the green square indicates the corrupted area. In practice, that region is set to zero as input to the network. }
    \label{fig:sp}
\end{figure*}

\section{Results}
% \subsection{Performance Measures}
We evaluate our model using three metrics: the \textit{MAE} between the  magnitude spectrogram and the ground-truth within the corrupted speech region; \textit{STOI} \cite{taal2011algorithm}, a speech intelligibility estimator; and \textit{PESQ} \cite{pesq}, a speech quality estimator. \textit{STOI} and \textit{PESQ} scores lie between -1 and 1, and -0.5 and 4.5, respectively. While lower \textit{MAE} scores corresponds to a lower reconstruction loss, for \textit{PESQ} and \textit{STOI}, the higher the better. 

Since it is not possible to use \textit{STOI} and \textit{PESQ} for signals shorter than a few hundreds ms \cite{taal2011algorithm,pesq}, we cannot use them only on the corrupted part. Therefore, we compute the scores for the whole signal. This lowers the sensitivity of the metrics, especially when inpainting short segments.
%As explained in  Sec. \ref{dataset}, we generate artificial corruptions in the innermost part of the excerpt, so that a non-silent part is always affected.

\subsection{Constrained Vocabulary Performance} \label{sec:consperf}
As our goal is to develop a system capable of dealing with corrupted segments of any duration, rather than training a system specifically for each gap length, in Table \ref{tab:results}, we report the overall performance of the model trained and evaluated in the \textit{Grid Corpus} for a distribution of segment durations that matches that of the training stage.  As it can be clearly seen, the proposed AV model is not only better than its AO counterpart, but it also outperforms the previous state-of-the-art AV model \cite{morrone2021audio}. 
%Notice that the \textit{PESQ} and \textit{STOI} values reported in \cite{morrone2021audio} are larger that the ones obtained in our experiments. As explained in  Sec. \ref{dataset}, we generate artificial corruptions in the innermost part of the audio excerpt, so that a non-silent part is always affected.

\begin{table}[th]
\caption{Performance scores averaged across the \textit{Grid} test set. Corrupted segment lengths sampled from a uniform distribution. The symbol  $\uparrow$ ($\downarrow$) indicates higher (lower) the better.}
\label{tab:results}
\centering
\begin{tabular}{lccc}
\toprule
                   & PESQ $\uparrow$ & STOI $\uparrow$ & MAE $\downarrow$   \\ \hline
Corrupted input    & 1.78 & 0.58 & 0.43 \\
Morrone et al. \cite{morrone2021audio}            & 1.98 & 0.79 & 0.39 \\
Proposed, audio-only   & 2.07 & 0.79 & 0.34 \\
Proposed, audio-visual & \textbf{2.21} & \textbf{0.84} & \textbf{0.31}\\
\bottomrule
\end{tabular}
%\caption{Performance scores averaged across the \textit{Grid} test set. Corrupted segment lengths sampled from a uniform distribution. The symbol  $\uparrow$ ($\downarrow$) indicates higher (lower) the better.}
%\label{tab:results}
\end{table}

\subsection{Performance vs Segment Duration} \label{sec:perdur}
From Table \ref{tab:results}, we can notice that the performance of the AV baseline,  \cite{morrone2021audio}, is worse than the proposed AO model. As AO methods are good at inpainting short gaps, we carried out an analysis of the performance of each model against the corrupted segment duration. The results are shown in  Fig. \ref{fig:galaxy}. 

Considering the \textit{MAE} values, we can see that they do not change significantly for segments larger than 800 ms. We hypothesize that the uncorrupted audio is used to determine the voice characteristics and  the speech continuity in the boundary of the corrupted segment, while the rest is purely generated from the visuals. That is why the AO model works well for short gaps, where the missing information can be inferred from the audio context, and fails to inpaint larger gaps. Besides, the relative \textit{MAE} between the reconstructed segments of 1600 ms and 160 ms (27\% for the AO model and around $10\%$  %$\sim10\%$ 
for the AV models) shows the effectiveness of the AV methods, as the \textit{MAE} degradation of the AO model is much higher. %than the one of models processing AV signals.

Analysing the results for each segment duration, the performance of the proposed AV and AO models, according to the estimated intelligibility (\textit{STOI}), is roughly similar when considering corrupted segments of 160 ms. 

On contrary, for corrupted segments of 400 ms, the proposed AV model is better at intelligibility and perceived quality. Nevertheless, the AO model is still very effective. In Fig. \ref{fig:sp} we can observe how the spectrogram predicted by the AO model is similar to that of the AV model, even the harmonics are better-defined than in the AV baseline's spectrogram. 

For corrupted segments of 800 ms and 1600 ms, the proposed AV model is the best. For such long gaps, the AO model is no longer capable of estimating the content of the sentence. It just generates a kind of mumbling, either as a consequence of inpainting the sample with certain energy bands that match the harmonics of the voice or as an attempt to mimic sentences learned from the dataset. If we consider \textit{PESQ}, we can see that, for segments of 1600 ms, the scores for the models tend to collapse to a single point. Our hypothesis is that, for such a long gap, the speech context is almost non-existent (see the first row of Fig.\ \ref{fig:sp}), therefore the task becomes close to speech reconstruction from silent videos \cite{michelsanti2021overview}, for which speech characteristics of unknown speakers, that are important for \textit{PESQ}, cannot be easily estimated using only the video information.

\subsection{Natural Vocabulary Performance}
In Sections \ref{sec:consperf} and \ref{sec:perdur} we evaluated the model in the \textit{Grid Corpus}, where the vocabulary is limited and unnatural. It is of our interest to study the performance of the model with real-world, in-the-wild data. In Table \ref{tab:voxresults}, we evaluate the model trained in the English subset of \textit{Voxceleb2}. %, described at \textcolor{red}{CITA}. 
Both, the baseline model from \cite{morrone2021audio} as well as the AO model did not converge when training in \textit{Voxceleb2}.
\begin{table}[th]
\caption{Performance scores averaged across \textit{Voxceleb2} test set. Corrupted segment lengths sampled from a uniform distribution.  The symbol  $\uparrow$ ($\downarrow$) indicates higher (lower) the better.}
\centering
\begin{tabular}{lccc}
\toprule
                   & PESQ $\uparrow$ & STOI $\uparrow$ & MAE $\downarrow$   \\ \hline
Corrupted input    & 1.37 & 0.43 & 0.56 \\
Proposed, audio-visual & \textbf{1.95} & \textbf{0.70} & \textbf{0.37}\\
\bottomrule
\end{tabular}
\label{tab:voxresults}
\end{table}

The results show how the proposed AV model is capable of generating meaningful speech on in-the-wild scenarios with unconstrained vocabulary, unlike the baseline and the AO model. This reflects the capabilities of AV models to synthesize speech in complex scenarios.
% \subsection{Limitations}

Some demos of reconstructed audio signals (both in \textit{Grid} and \textit{Voxceleb2} test samples) are available at \url{https://ipcv.github.io/avsi/}.

\section{Conclusions and Future Work}
This paper presented a new state-of-the-art AVSI model that can inpaint long gaps, up to 1600 ms, for unseen-unheard speakers.
We tested our model in the \textit{Grid Corpus} \cite{grid} and showed that it outperforms its audio-only counterpart for gaps larger than 160 ms, and
the  previous state-of-the-art approach. In addition, we showed that the visual features extracted from the AV-HuBERT network encode enough information to guide the inpainting process. Besides, we showed our model can inpaint natural, unconstrained speech in in-the-wild scenarios (\textit{Voxceleb2} dataset \cite{voxceleb}).
One of the limitations of the proposed and the existing AVSI approaches is that the mapping between phonemes and visemes is not bijective, namely, a single viseme may correspond to many phonemes \cite{fisher1968confusions}. For example, the sentences \textit{``elephant juice"} and \textit{``I love you"} share the same visemes. To overcome this limitation, additional information may be included, such us context information about the scenario or language models.
% \vfill\pagebreak 
% \newpage

\section{Acknowledgements}
This work has been supported by MICINN/FEDER UE project
PID2021-127643NB-I00 and FPI grant PRE2018-083920. %We acknowledge NVIDIA Corporation for the donation of GPUs used for the experiments

% References should be produced using the bibtex program from suitable
% BiBTeX files (here: strings, refs, manuals). The IEEEbib.bst bibliography
% style file from IEEE produces unsorted bibliography list.
% -------------------------------------------------------------------------
\bibliographystyle{IEEEtran}
\bibliography{refs}

% Generated by IEEEtran.bst, version: 1.14 (2015/08/26)
\begin{thebibliography}{10}
\providecommand{\url}[1]{#1}
\csname url@samestyle\endcsname
\providecommand{\newblock}{\relax}
\providecommand{\bibinfo}[2]{#2}
\providecommand{\BIBentrySTDinterwordspacing}{\spaceskip=0pt\relax}
\providecommand{\BIBentryALTinterwordstretchfactor}{4}
\providecommand{\BIBentryALTinterwordspacing}{\spaceskip=\fontdimen2\font plus
\BIBentryALTinterwordstretchfactor\fontdimen3\font minus
  \fontdimen4\font\relax}
\providecommand{\BIBforeignlanguage}[2]{{%
\expandafter\ifx\csname l@#1\endcsname\relax
\typeout{** WARNING: IEEEtran.bst: No hyphenation pattern has been}%
\typeout{** loaded for the language `#1'. Using the pattern for}%
\typeout{** the default language instead.}%
\else
\language=\csname l@#1\endcsname
\fi
#2}}
\providecommand{\BIBdecl}{\relax}
\BIBdecl

\bibitem{6020748}
A.~Adler, V.~Emiya, M.~G. Jafari, M.~Elad, R.~Gribonval, and M.~D. Plumbley,
  ``Audio inpainting,'' \emph{IEEE Transactions on Audio, Speech, and Language
  Processing}, vol.~20, no.~3, pp. 922--932, 2012.

\bibitem{ebner2020audio}
P.~P. Ebner and A.~Eltelt, ``Audio inpainting with generative adversarial
  network,'' \emph{arXiv preprint arXiv:2003.07704}, 2020.

\bibitem{chang2019deep}
Y.-L. Chang, K.-Y. Lee, P.-Y. Wu, H.-y. Lee, and W.~Hsu, ``Deep long audio
  inpainting,'' \emph{arXiv}, 2019.

\bibitem{marafioti2019context}
A.~Marafioti, N.~Perraudin, N.~Holighaus, and P.~Majdak, ``A context encoder
  for audio inpainting,'' \emph{IEEE/ACM Transactions on Audio, Speech, and
  Language Processing}, vol.~27, no.~12, pp. 2362--2372, 2019.

\bibitem{kegler2019deep}
M.~Kegler, P.~Beckmann, and M.~Cernak, ``Deep speech inpainting of
  time-frequency masks,'' \emph{Interspeech}, 2020.

\bibitem{speechpainter}
Z.~Borsos, M.~Sharifi, and M.~Tagliasacchi, ``Speechpainter: Text-conditioned
  speech inpainting,'' \emph{Interspeech}, 2022.

\bibitem{morrone2021audio}
G.~Morrone, D.~Michelsanti, Z.-H. Tan, and J.~Jensen, ``Audio-visual speech
  inpainting with deep learning,'' in \emph{IEEE International Conference on
  Acoustics, Speech and Signal Processing (ICASSP)}, 2021, pp. 6653--6657.

\bibitem{grid}
M.~Cooke, J.~Barker, S.~Cunningham, and X.~Shao, ``An audio-visual corpus for
  speech perception and automatic speech recognition,'' \emph{The Journal of
  the Acoustical Society of America}, vol. 120, no.~5, pp. 2421--2424, 2006.

\bibitem{voxceleb}
J.~S. Chung, A.~Nagrani, and A.~Zisserman, ``Voxceleb2: Deep speaker
  recognition,'' in \emph{Interspeech}, 2018.

\bibitem{avhubert}
B.~Shi, W.-N. Hsu, K.~Lakhotia, and A.~Mohamed, ``Learning audio-visual speech
  representation by masked multimodal cluster prediction,'' \emph{ICLR}, 2022.

\bibitem{paulino2020paco}
I.~R. Paulino and I.~Hounie, ``Paco and paco-dct: Patch consensus and its
  application to inpainting,'' in \emph{IEEE International Conference on
  Acoustics, Speech and Signal Processing (ICASSP)}, 2020, pp. 5775--5779.

\bibitem{resnet}
K.~He, X.~Zhang, S.~Ren, and J.~Sun, ``Deep residual learning for image
  recognition,'' in \emph{2016 IEEE Conference on Computer Vision and Pattern
  Recognition (CVPR)}, 2016, pp. 770--778.

\bibitem{rahimi2022reading}
A.~Rahimi, T.~Afouras, and A.~Zisserman, ``Reading to listen at the cocktail
  party: Multi-modal speech separation,'' in \emph{Proceedings of the IEEE/CVF
  Conference on Computer Vision and Pattern Recognition}, 2022, pp.
  10\,493--10\,502.

\bibitem{kadandale2022vocalist}
V.~S. Kadandale, J.~F. Montesinos, and G.~Haro, ``Vocalist: An audio-visual
  synchronisation model for lips and voices,'' \emph{Interspeech}, 2022.

\bibitem{chen2021audio}
H.~Chen, W.~Xie, T.~Afouras, A.~Nagrani, A.~Vedaldi, and A.~Zisserman,
  ``Audio-visual synchronisation in the wild,'' \emph{32nd British Machine
  Vision Conference (BMVC)}, 2021.

\bibitem{montesinos2022vovit}
J.~F. Montesinos, V.~S. Kadandale, and G.~Haro, ``Vovit: Low latency
  graph-based audio-visual voice sseparation transformer,'' in \emph{European
  Conference on Computer Vision (ECCV)}, 2022.

\bibitem{lws}
J.~Le~Roux, H.~Kameoka, N.~Ono, and S.~Sagayama, ``Fast signal reconstruction
  from magnitude stft spectrogram based on spectrogram consistency,'' in
  \emph{Proc. DAFx}, vol.~10, 2010, pp. 397--403.

\bibitem{gelu}
D.~Hendrycks and K.~Gimpel, ``Gaussian error linear units (gelus),''
  \emph{arXiv preprint arXiv:1606.08415}, 2016.

\bibitem{elu}
D.~Clevert, T.~Unterthiner, and S.~Hochreiter, ``Fast and accurate deep network
  learning by exponential linear units (elus),'' in \emph{4th International
  Conference on Learning Representations, {ICLR}}, 2016.

\bibitem{taal2011algorithm}
C.~H. Taal, R.~C. Hendriks, R.~Heusdens, and J.~Jensen, ``An algorithm for
  intelligibility prediction of time--frequency weighted noisy speech,''
  \emph{IEEE Transactions on Audio, Speech, and Language Processing}, vol.~19,
  no.~7, pp. 2125--2136, 2011.

\bibitem{pesq}
A.~W. Rix, J.~G. Beerends, M.~P. Hollier, and A.~P. Hekstra, ``Perceptual
  evaluation of speech quality (pesq)-a new method for speech quality
  assessment of telephone networks and codecs,'' in \emph{IEEE International
  Conference on Acoustics, Speech and Signal Processing (ICASSP)}, vol.~2,
  2001, pp. 749--752.

\bibitem{michelsanti2021overview}
D.~Michelsanti, Z.-H. Tan, S.-X. Zhang, Y.~Xu, M.~Yu, D.~Yu, and J.~Jensen,
  ``An overview of deep-learning-based audio-visual speech enhancement and
  separation,'' \emph{IEEE/ACM Trans. on Audio, Speech, and Language
  Processing}, vol.~29, pp. 1368--1396, 2021.

\bibitem{fisher1968confusions}
C.~G. Fisher, ``Confusions among visually perceived consonants,'' \emph{Journal
  of speech and hearing research}, vol.~11, no.~4, pp. 796--804, 1968.

\end{thebibliography}

\end{document}